\documentclass[doublecol,linenumbers]{epl2}

\usepackage{graphicx}
\usepackage{subfigure}
\usepackage{amsthm}
\usepackage{amsmath}
\usepackage{amssymb}
\usepackage{verbatim}
\usepackage{dcolumn}
\usepackage{bm}
\usepackage{epsf}
\usepackage{color}
\usepackage[colorlinks=true,citecolor=blue,linkcolor=blue,urlcolor=blue]{hyperref}%
\usepackage{xcolor}
\usepackage{dsfont}
\newcommand{\bra}[1]{\left\langle #1\right|}
\newcommand{\ket}[1]{\left|#1\right\rangle}
\newcommand{\braket}[2]{\left\langle #1|#2\right\rangle}

\newcommand{\tr}[1]{\mathrm{tr}\left\{#1\right\}}

\newcommand{\la}{\left\langle}
\newcommand{\ra}{\right\rangle}
\newcommand{\pd}{\partial}

\newcommand{\e}[1]{\exp{\left(#1\right)}}
\newcommand{\lo}[1]{\ln{\left(#1\right)}}

\newcommand{\bla}{bla\\bla\\bla\\bla\\bla}
\newcommand{\mb}[1]{\mbox{\boldmath$#1$}}
\newcommand{\mc}[1]{\mathcal{#1}}

\newcommand{\mf}[1]{\mathfrak{#1}}
\newcommand{\mrm}[1]{\mathrm{#1}}

\DeclareMathOperator*{\sumint}{%
\mathchoice%
  {\ooalign{$\displaystyle\sum$\cr\hidewidth$\displaystyle\int$\hidewidth\cr}}
  {\ooalign{\raisebox{.14\height}{\scalebox{.7}{$\textstyle\sum$}}\cr\hidewidth$\textstyle\int$\hidewidth\cr}}
  {\ooalign{\raisebox{.2\height}{\scalebox{.6}{$\scriptstyle\sum$}}\cr$\scriptstyle\int$\cr}}
  {\ooalign{\raisebox{.2\height}{\scalebox{.6}{$\scriptstyle\sum$}}\cr$\scriptstyle\int$\cr}}
}

\title{Energetic cost of Hamiltonian quantum gates}

\author{Sebastian Deffner\inst{1,2} \thanks{E-mail: \email{deffner@umbc.edu}}}
\shortauthor{Sebastian Deffner}
\institute{
  \inst{1} Department of Physics, University of Maryland, Baltimore County, Baltimore, MD 21250, USA\\
  \inst{2} Instituto de F\'isica `Gleb Wataghin', Universidade Estadual de Campinas, 13083-859, Campinas, S\~{a}o Paulo, Brazil}


\date{\today}

\abstract{Landauer's principle laid the main foundation for the development of modern thermodynamics of information. However, in its original inception the principle relies on semiformal arguments and dissipative dynamics. Hence, if and how Landauer's principle applies to unitary quantum computing is less than obvious.  Here, we prove an inequality bounding the change of Shannon information encoded in the logical quantum states by quantifying the energetic cost of Hamiltonian gate operations. The utility of this bound is demonstrated by outlining how it can be applied to identify energetically optimal quantum gates in theory and experiment. The analysis is concluded by discussing the energetic cost of quantum error correcting codes with non-interacting qubits, such as Shor's code.}

\begin{document}

\maketitle

\section{Introduction}

Introductory textbooks often present thermodynamics as a phenomenological framework, whose processes are driven by the exchange of heat and work \cite{Callen1985}. However, ever since the birth of Maxwell's demon \cite{Leff2014},  describing and understanding the properties of \emph{information} has been an integral part of thermodynamics \cite{Bekenstein1981,Pendry1983,Landauer1987,Bekenstein1988,Zurek1989,Zurek2018,Alicki2019}.  Nevertheless, the \emph{thermodynamics of information} has only rather recently become a prominent area of research, which has been attracting more and more attention \cite{Mandal2012,Mandal2013,Deffner2013PRE,Deffner2013PRX,Parrondo2015,Strasberg2017,Safranek2018,Stevens2019,Miller2020,Touil2020,Touil2021}. This development arose out of \emph{stochastic thermodynamics} \cite{Seifert2008,Seifert2012,Ciliberto2017,Wolpert2019}, and its application to processes with feedback \cite{Sagawa2008,Sagawa2010}.

A seminal result that laid the foundation for the theory is Landauer's principle \cite{Landauer1961,Landauer1991}.  As a statement of the second law,  it is typically written as,
\begin{equation}
\label{eq:landauer}
\la W\ra \geq k_B T\,\lo{2}\,,
\end{equation}
which states that in order to erase one bit of classical information at temperature $T$, at least $k_B T\,\lo{2}$ of thermodynamic work,  $\la W\ra$, are required.  Although the original derivation \cite{Landauer1961} can be considered semiformal at best, the impact of Eq.~\eqref{eq:landauer} on the development of computing hardware can hardly be underestimated \cite{Bennet2003}. Nevertheless,  while recent experiments in classical \cite{Berut2012,Jun2014,Hong2016} as well as quantum systems \cite{Peterson2016,Gaudenzi2018} verified the validity of Landauer's principle,  they also demonstrated its shortcomings and inadequacy when applied to reversible  \cite{Bennet2003} or quantum computing \cite{Nielsen2010}. 

In particular, quantum computing \cite{Nielsen2010} is designed from unitary dynamics, which describes systems that are thermally isolated from the environment.  Any external noise inevitably thwarts so-called quantum advantage \cite{Zurek2003,Sanders2017,Schlosshauer2019},  as it harms the delicate nature of quantum states. Yet, Landauer's principle \eqref{eq:landauer} and its generalizations to open systems \cite{Hilt2011,Reeb2014,Lorenzo2015,Faist2015,Goold2015,Man2019,Ayhin2020} are specifically formulated for dissipative dynamics at finite temperature.  Thus, some recent efforts have sought to remedy this issue, by either analyzing the energetics of measurements and quantum operations directly \cite{Gea2002,Bedingham2016,Cimini2020}, or by generalizing notions of stochastic thermodynamics to zero temperature \cite{Timpanaro2020}. For a more comprehensive review of recent developments, we refer to the literature \cite{Goold2016,Deffner2019book}.

The present analysis is dedicated to an alternative treatment, and the main goal is to quantify the energetic cost of single gate operations in unitary quantum computing. To this end, we derive an upper bound on the change of Shannon information encoded in the marginal, logical states of a quantum system, that evolves under a Hamiltonian gate operation \cite{Santos2017}. We find that this upper bound is given by the norm of the Hamiltonian, which has been proposed in the literature to quantify the energetic cost of quantum control protocols \cite{Zheng2016,Campbell2017,Abah2019}.  Thus, as a main result, we obtain an inequality relating the amount of processed information with the energetic cost of the operation -- a generalized Landauer's principle for quantum computing.  The versatility of this bound is demonstrated for qubit reset, i.e., for a Hadamard gate. The discussion is concluded with a few remarks on the experimental relevance of the study, and with remarks on its utility in assessing the energetic overhead of quantum error correcting codes.

\section{Preliminaries: logical states and marginal information}

In quantum statistical physics as well as in quantum information theory \cite{Nielsen2010} the amount of information stored in quantum state is quantified by the von Neuman entropy,
\begin{equation}
S_\mrm{vN}=-\tr{\rho \lo{\rho}}\,,
\end{equation}
where $\rho=\sumint_\Gamma p(\Gamma) \ket{\psi(\Gamma)}\bra{\psi(\Gamma)}$ is the density operator. The issue now arises from the fact that $S_\mrm{vN}$ is invariant under unitary transformations. Hence, the informational content of a quantum state does not appear to change under a quantum computation realized with unitary gates.

However, it also has been recognized that in assessing the thermodynamics of a computation not only the bare information content, but also the algorithmic complexity, the self-entropy,  i.e., the entropy carried in a single logical state, etc. need to be considered \cite{Zurek1989,Boyd2016NJP,Wolpert2019}. The concept can be easily illustrated for a simple operation on a qubit. 

Imagine a two-level system that is initially prepared in an even superposition of logical states $\ket{0}$ and $\ket{1}$,
\begin{equation}
\ket{\psi_\mrm{in}}= \frac{1}{\sqrt{2}}\left(\ket{0}+\ket{1}\right)\,.
\end{equation}
Then a Hadamard gate is applied, which we write as
\begin{equation}
\label{eq:hadamard}
\mc{H}=\frac{1}{\sqrt{2}}\left(\ket{0}\bra{0}+\ket{1}\bra{0}+\ket{0}\bra{1}-\ket{1}\bra{1}\right)\,.
\end{equation}
In the present example,  the Hadamard gate  performs nothing but a qubit ``reset''.  The ``output'' state then becomes
\begin{equation}
\ket{\psi_\mrm{out}}=\mc{H}  \ket{\psi_i}=\ket{0}\,.
\end{equation}
Since $\mc{H}$ is unitary, the von Neumann entropy, i.e., the informational content of the pure state has not changed and remains to be zero.

However, in the input state the probability to observe the logical state $\ket{0}$ is $1/2$, and in the output state it is $1$.  In other words, the distribution over the logical states has changed. This can be described by the Shannon information, $\mc{S}$, of the marginal distribution, that is the entropy of the quantum state that has been measured in the logical basis,
\begin{equation}
\label{eq:shannon}
\mc{S}= -\sum_{n\in\{0,1\}}  p_n\lo{p_n}\,,
\end{equation}
where $p_n=|\braket{n}{\psi}|^2$.

For this simple example, the change of the Shannon information then becomes,
\begin{equation}
\Delta\mc{S}=\mc{S}_\mrm{out}-\mc{S}_\mrm{in}=\lo{2}\,.
\end{equation}
The natural question arises, how to quantify the energetic cost for this change of information. In complete analogy to previous considerations \cite{Zurek1989,Boyd2016,Deffner2013PRX,Wolpert2019} it appears obvious that such a change of information cannot be ``for free'', and that some generalized version of Landauer's principle \eqref{eq:landauer} must apply.

\section{Cost of Hamiltonian gates} 

We now proceed to derive such a generalized Landauer's principle for unitary dynamics. To this end, we consider pure quantum states, $\rho(t)=\ket{\psi(t)}\bra{\psi(t)}$, that evolve under the time-dependent Schr\"odinger equation
\begin{equation}
\label{eq:schrodinger}
 i\hbar\,\pd_t \ket{\psi(t)}=H(t) \ket{\psi(t)}\,.
\end{equation}
Only rather recently, it was shown how time-dependent Hamiltonians can be constructed by inverse engineering such that the evolution under Eq.~\eqref{eq:schrodinger} is equivalent to any desired quantum gate \cite{Santos2017}.  Hence, for our present purposes it will be sufficient to work with arbitrary, time-dependent Hamiltonians.  

In the following, we will bound the change of the Shannon information  \eqref{eq:shannon} under the driven dynamics \eqref{eq:schrodinger}.  In the above example, we consider only a binary logical basis. However,  the following analysis is more general,  since we formulate the derivation for any arbitrary marginal, which, e.g., includes the diagonal entropy \cite{Polkovnikov2011}.  We also note that the following arguments are somewhat reminiscent of our previous work \cite{Deffner2020PRR}, yet the final result and conclusions are conceptually markedly different.

\subsection{Bound on marginal entropy}

We start by considering the magnitude of the change of Shannon information \eqref{eq:shannon} under the time-dependent Schr\"odinger equation \eqref{eq:schrodinger}, which can be bounded from above with the triangle inequality,
\begin{equation}
\label{eq:inequality_1}
|\Delta\mc{S}|\leq \int_0^\tau \upd t\, |\dot{\mc{S}}(t)|\,.
\end{equation}
Due to normalization, the rate of change of $\mc{S}$ simply reads, $\dot{\mc{S}}(t)=-\sum_n \dot{p}_n \lo{p_n}$.  Hence, we continue by inspecting $\dot{p}_n(t)$. Noting that
\begin{equation}
\dot{p}_n(t)=\pd_t \braket{\psi(t)}{n}\braket{n}{\psi(t)}
\end{equation}
we immediately have
\begin{equation}
\label{eq:inequality_1a}
\dot{p}_n(t)\leq 2 \hbar\, \left|\bra{n}H(t)\ket{\psi(t)}\braket{\psi(t)}{n}\right|\,.
\end{equation}
It is then convenient to write Eq.~\eqref{eq:inequality_1a} as
\begin{equation}
\dot{p}_n(t)\leq 2 \hbar\, \left|\tr{H\,\ket{n}\braket{n}{\psi}\bra{\psi}}\right|
\end{equation}
where we suppressed the explicit time-dependence to avoid clutter, and which can be further bounded by the H\"older inequality \cite{Baumgartner2011}. This theorem states that
\begin{equation}
\label{eq:holder}
\left| \tr{A\,B^\dagger}\right|\leq \left(\tr{|A|^p}\right)^{1/p}\,\left(\tr{|B|^q}\right)^{1/q}
\end{equation}
for all non-negative $p$ and $q$ with $1/p+1/q=1$.  The right side of Eq.~\eqref{eq:holder} is given by the product of the Schatten-$p$ and Schatten-$q$ norms of the operators $A$ and $B$ respectively, which can be expressed in terms of the singular values. For instance, the Schatten-1 norm, i.e., the trace norm can be written as $\tr{|A|}=\sum_\nu \sigma_\nu$, where $\sigma_n$ are the singular values of $A$.

Now, choosing $q=1$ and $p=\infty$ for $A=H(t)$ and $B=\ket{n}\braket{n}{\psi}\bra{\psi}$, our case becomes particularly simple. Note that the operator $\ket{n}\braket{n}{\psi}\bra{\psi}$ has only one singular value that is different from zero, namely $\sqrt{p_n}=|\braket{\psi}{n}|$.  Hence, we immediately obtain
\begin{equation}
\label{eq:inequality_2}
\dot{p}_n(t)\leq 2 \hbar\, ||H(t)||\, \sqrt{p_n}\,,
\end{equation}
where $||H(t)||$ is the operator norm, i.e., the largest singular value of the time-dependent Hamiltonian.

Using Eq.~\eqref{eq:inequality_2}, the change of Shannon information \eqref{eq:inequality_1} can now be bounded by
\begin{equation}
\label{eq:first_result}
|\Delta\mc{S}|\leq 2\hbar \int_0^\tau \upd t\, ||H(t)||\,\mf{S}(t)
\end{equation}
where we introduce $\mf{S}(t)\equiv-\sum_n \sqrt{p_n}\, \log{p_n}\geq 0$.  Equation~\eqref{eq:first_result} constitutes our first main result.  The change of Shannon information of the marginal (post-measurement) distribution is upper bounded by the time-convolution of the norm of the Hamiltonian and $\mf{S}(t)$. While Eq.~\eqref{eq:first_result} is mathematically simple and appealing, the physical interpretation is not quite as transparent as one would desire.

Therefore, we continue by further bounding $\mf{S}(t)$ by its maximum that can be supported by the quantum system. For $d$-dimensional Hilbert spaces, with $d<\infty$, we can write
\begin{equation}
\mf{S}(t)\leq \sqrt{d}\,\lo{d}\,,
\end{equation}
which follows from similar arguments as maximizing the Shannon information. Further introducing the information in units of bits $\mc{I}\equiv\mc{S}/\lo{2}$, we finally obtain
\begin{equation}
\label{eq:qm_landauer}
|\Delta\mc{I}|\leq 2\hbar\, \sqrt{d}\,\log_2{(d)}\,\int_0^\tau \upd t\, ||H(t)||\,.
\end{equation}
Quite remarkably,  the time integrated norm of the Hamiltonian has been discussed in the literature to quantify the energetic cost of quantum control protocols \cite{Santos2015,Zheng2016,Campbell2017,Santos2017JPA,Hu2018,Abah2019}. For instance, for spin systems this cost can be interpreted as the average power expended by magnetic control fields. Further comparing Eq.~\eqref{eq:qm_landauer} with the original Landauer's principle \eqref{eq:landauer}, we immediately observe that for single bit operations, $d=2$,  we have indeed achieved a generalized Landauer's principle for Hamiltonian gates.

Finally,  Eq.~\eqref{eq:qm_landauer} can also be generalized to infinite-dimensional Hilbert spaces with bounded energy, $E_0$.  In that case, the maximal Shannon information is given by the Gibbs entropy with effective, inverse  temperature $\beta$, such that the average energy is equal to $E_0$. See also Ref.~\cite{Deffner2020PRR} for a related discussion.

\subsection{Reset of a logical qubit}

We conclude this section by returning to the aforementioned example of resetting a qubit. It has been shown by Santos \cite{Santos2017} that the Hadamard gate $\mc{H}$ \eqref{eq:hadamard} can be implemented through the time-dependent Hamiltonian
\begin{equation}
H_\mc{H}(t)=\frac{\dot{\varphi}(t)}{2\sqrt{2}\,\hbar}\, \left(\sigma_x+\sigma_z\right)
\end{equation}
where $\varphi(t)$ is an arbitrary function fulfilling the boundary conditions $\varphi(0)=0$ and $\varphi(\tau)=\pi$. Note that physically $\dot{\varphi}(t)$ is nothing but the strength of the magnetic field, $\mb{B}(t) \propto \dot{\varphi}(t)\, (1,0,1)$. One easily convinces oneself that
\begin{equation}
\mc{H}=\mc{T}_> \e{\int_0^\tau \upd t\, H_\mc{H}(t)}\,,
\end{equation}
where  $\mc{T}_>$ denotes time-ordering.

Realizing now $||\sigma_x+\sigma_z||=\sqrt{2}$, it is easy to see that the generalized Landauer's principle \eqref{eq:qm_landauer} becomes
\begin{equation}
\label{eq:H_landauer}
 |\Delta\mc{I}|\leq \sqrt{2}\,\int_0^\tau \upd t\, |\dot{\varphi}(t)|\,.
\end{equation}
Hence, we immediately observe that the energetic cost for resetting a qubit is simply given by the magnitude of the magnetic field employed to realize the quantum gate. Moreover, it then becomes a problem of optimal control theory to design energetically optimal Hadamard gates, which are characterized by minimizing the time-integrated magnitude of the magnetic field \cite{Deffner2014JPB}.  Interestingly, this is a problem that has already been studied extensively in the literature on thermodynamic control \cite{Zulkowski2015,Acconica2015,Deffner2020EPL,Saira2020}.

The full solution of the optimal control problem for finding Hamiltonian quantum gates with minimal (average) intensity is beyond the scope of the present discussion. However, let us briefly outline how such an analysis would work. In Ref.~\cite{Santos2017} it was alluded to the fact that a Hamiltonian Hadamard gate can, indeed, be realized in experiments with the linear protocol,
\begin{equation}
\label{eq:linear}
\varphi_0(t)=\pi t/\tau\,.
\end{equation}
Optimal protocols can then be found by considering, e.g., a Fourier ansatz
\begin{equation}
\varphi(t)=\varphi_0(t) +\sum_k A_k \sin{\left(2\pi \,k\,t/\tau \right)}\,.
\end{equation}
Also see Ref.~\cite{Abah2019} for similar considerations in the context of shortcuts to adiabaticity.  Note, however, that is has been shown in the literature \cite{Bonanca2018} that more involved series expansions,  as for instance in terms of the Chebyshev polynomials have better performance.  In any case, the optimization problem then reduces to finding the set of coefficients $\{A_k\}_k$ such that the right side of the inequality \eqref{eq:H_landauer} becomes minimal. However, it is worth emphasizing that any polynomial ansatz must fulfill the boundary conditions, $\varphi(0)=0$ and $\varphi(\tau)=\pi$, which does limit the choice of possible driving protocols.

\begin{figure}
\includegraphics[width=.48\textwidth]{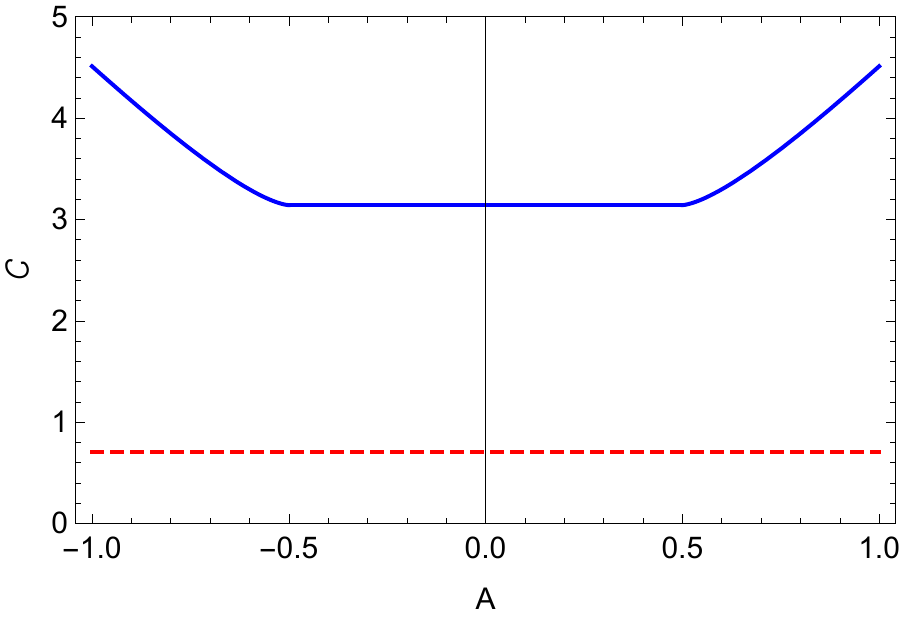}
\caption{\label{fig1} Illustration of the generalized Landauer's principle for qubit reset \eqref{eq:H_landauer} for the linear protocol plus one Fourier mode \eqref{eq:prot}. Left side of Eq.~\eqref{eq:H_landauer}, $ |\Delta\mc{I}|/\sqrt{2}=1/\sqrt{2}$, as red,  dashed line, and $\mc{C}\equiv \int_0^\tau \upd t\, |\dot{\varphi}(t)|$ depicted as blue,  solid line.}
\end{figure}

In the simplest case,  only one Fourier component in addition to the linear protocol is available, and we have 
\begin{equation}
\label{eq:prot}
\varphi_1(t)= \pi t/\tau+A \sin{\left(2\pi \,t/\tau \right)}\,.
\end{equation}
It is then a simple exercise to show  with $\mc{C}\equiv \int_0^\tau \upd t\, |\dot{\varphi}(t)|$ that
\begin{equation}
\mc{C}=\begin{cases} 2\left[-\mrm{arccsc}{\left(2 A\right)+\sqrt{4 A^2-1}}\right],&\text{for}\, A\leq -1/2\\
\pi, &\text{for}\, A\in (-1/2,1,2)\\
2\left[\mrm{arccsc}{\left(2 A\right)+\sqrt{4 A^2-1}}\right], &\text{for}\, A\geq 1/2\,.\end{cases}
\end{equation}
Hence, we conclude that if only the linear protocol plus a single Fourier mode is available, then the linear protocol \eqref{eq:linear} is energetically optimal. See also Fig.~\ref{fig1} for an illustration of this finding. Finally we note that while Eq.~\eqref{eq:H_landauer} is of mathematically simple and appealing form, the bound  is not particularly tight. For this simple example, the minimum of the right side turns out to be $\sqrt{2} \pi$, whereas the left side is only $1$.  Significantly tighter bounds can be obtained by directly working with Eq.~\eqref{eq:first_result}. 

\section{Applications and consequences}

We conclude the analysis with a discussion of potential applications and consequences of the generalized Landauer's principle \eqref{eq:qm_landauer} for quantum computing.  Broadly speaking, there are two avenues of research \cite{Sanders2017} that appear plausible,  namely the experimental realization of logical qubits and the implementation of quantum error correcting codes.

\subsection{Energetics of experimental gate operations}

To date several computational paradigms have been proposed, of which quantum circuits or gate based quantum computing \cite{Nielsen2010}, adiabatic quantum computing based on quantum annealing \cite{Das2008}, and cluster state computing \cite{Nielsen2006} have received some prominence.  In addition,  possible hardware for quantum computing has been developed in, e.g., quantum optics \cite{Kok2007}, ion traps \cite{Bruzewicz2019}, and solid state systems \cite{Kjaergaard2020}.  Yet, independent of the computational paradigm and in any experimental platform the actual processing of information is facilitated by applying time-dependent, external fields. Hence, Eq.~\eqref{eq:qm_landauer} does apply to any version of a quantum computer.

\subsection{Superconducting qubits} As an illustrative example, consider transmon \cite{Koch2007}, charge \cite{Schreier2008}, or flux qubits \cite{Orlando1999}, which are all essentially based on a Cooper pair box \cite{Bouchiat1998},  and which are the basis of many currently available systems, such as IBM's Q Experience and the D-Wave machine \cite{Sanders2017}. The Hamiltonian of such a Cooper pair box reads \cite{Bouchiat1998}
\begin{equation}
H=-\frac{1}{2}\,\left(E_J\,\sigma_x+E\,\sigma _z\right)
\end{equation}
where $E=E_C\left(1-2 n_g\right)$ and $n_g=C_g/(2 e) U$ is the dimensionless gate voltage at capacitance $C_g$.  Further, $E_J$ is the Josephson energy, which is proportional to the area of the tunnel junction. Any quantum gate can then be implemented by applying an external magnetic field \cite{Santos2017}. Obviously, it is desirable to work with the magnetic fields that have -- on average -- the lowest intensity, to avoid excessive dissipation and decoherence. Equation~\eqref{eq:qm_landauer} then provides a simple tool, to determine the optimal fields with the minimally required intensity to realize the required quantum gate.

\subsection{Cost of quantum error correction}

It has been recognized that any reliable quantum computer will necessitate the implementation of \emph{quantum error correcting algorithms} \cite{Nielsen2010}. Loosely speaking, any such algorithm encodes logical quantum states in several physical states that can be controlled separately and in parallel.  Hence, the logical quantum states can be made resilient against the effects of noise, such as decoherence and dissipation. 

Reliable \emph{classical} computing is typically implemented by repetition codes.  \cite{Loepp2006}.  In essence, this just means that every single bit operation is performed $N$ times independently, and the outcome is determined from ``majority votes''.  Landauer's principle \eqref{eq:landauer} then quantifies the thermodynamic cost that arises from the ``physical overhead" of the error correcting code.  If the logical bit is encoded in $N$ physical bits, we immediately have that the total thermodynamic cost is simply given by $N$ times the work to erase a single bit, i.e., $N \times k_B T \ln(2)$.  The obvious question is if and how this argument carries over to quantum error correcting codes. To date, a plethora of algorithms has been proposed as e.g.,  Shor's code \cite{Shor1995}, topological codes \cite{Kitaev2006}, stabilizer codes \cite{Gottesman1997}, or entanglement-assisted schemes \cite{Brun2006}.  The generalized Landauer's principle \eqref{eq:qm_landauer} can then be exploited to rank these various codes according to their energetic cost.  

The analysis becomes particularly simple for algorithms that rely on \emph{non-interacting qubits}. Thus, the total Hamiltonian can be written as a sum of identical, single qubit Hamiltonians, $H(t)=\sum_{n=1}^N H_1(t)$.  If a logical qubit is encoded in $N$ physical qubits, Eq.~\eqref{eq:qm_landauer} can be written as
\begin{equation} 
\label{eq:N_landauer}
|\Delta\mc{I}|\leq 2\hbar\, \sqrt{d}\,\log_2{(d)}\,N\,\int_0^\tau \upd t\, ||H_1(t)||\,.
\end{equation}
Equation~\eqref{eq:N_landauer} follows from the linearity of the norm, since the $H_i(t)$ live only on the  Hilbert space of the $i$th qubit. Note that generally $\mc{I}\neq N \mc{I}_1$,  since the $N$ qubits are correlated in the logical basis.

\subsection{Shor's code} Arguably the most prominent quantum error correcting code was proposed by Shor \cite{Shor1995}. This algorithm is a generalization of classical repetition codes, and protects a single qubit against any arbitrary error, which has been demonstrated  in several experiments, see for instance Ref.~\cite{Briegel2000,Reed2012,Bell2014}.

In this scheme, the logical qubit is encoded in 9 physical qubits according to
\begin{equation}
\ket{0}\rightarrow\frac{\left(\ket{000}+\ket{111}\right)^3}{2\sqrt{2}}\quad\mrm{and}\quad \ket{1}\rightarrow\frac{\left(\ket{000}-\ket{111}\right)^3}{2\sqrt{2}}\,.
\end{equation}
Correspondingly,  Eq.~\eqref{eq:N_landauer} predicts that implementing Shor's code is 9 times as expensive as a single qubit operation.  Thus, the natural question arises whether there is a quantum error correcting code, that is less or even the least expensive. 

\subsection{Perfect quantum error correcting code} It was shown in Ref.~\cite{Laflamme1996} that the minimal quantum error correcting code requires only 5 physical qubits. Logical qubits are encoded (up to normalization) according to
\begin{equation}
\begin{split}
\ket{0}\rightarrow &-\ket{00000}+\ket{01111}-\ket{10011}+\ket{11100}\\
& +\ket{00110}+\ket{01001}+\ket{10101}+\ket{11010}
\end{split}
\end{equation}
and
\begin{equation}
\begin{split}
\ket{1}\rightarrow &-\ket{11111}+\ket{10000}+\ket{01100}-\ket{00011}\\
& +\ket{11001}+\ket{10110}-\ket{01010}-\ket{00101}\,.
\end{split}
\end{equation}
While the physical motivation and interpretation of this algorithm is somewhat obscure \cite{Laflamme1996}, it has been shown that the \emph{perfect quantum error correcting code} is, indeed, minimal and that it protects the logical qubits against any type of environmental noise.  Hence, the minimal energetic cost according to Eq.~\eqref{eq:N_landauer} of protecting single qubits is 5 times the cost of a single qubit operation.

\subsection{Quantum error correction with interactions} This simple, linear scaling does not hold for more intricate quantum error correcting codes that involve ``energy penalties''.  Prominent examples include the toric code,  which is a topological algorithm \cite{Kitaev2006}, and error correction in quantum annealers \cite{Sarovar2013,Young2013,Pudenz2015,Pastawski2016,Vinci2018}.  However, a thorough analysis of such algorithms is beyond the scope of the present discussion, and is thus postponed to future work.

\section{Concluding Remarks}

In the present analysis, we have derived a generalized Landauer's principle that can be used to quantify the energetic cost of Hamiltonian quantum gates. Remarkably, this novel bound holds for purely Hamiltonian dynamics, and hence does not rely on dissipative dynamics.  The utility of this bound has been demonstrated by alluding to the optimal control problem of finding Hamiltonian quantum gates with minimal intensity,  experimental considerations, and by assessing and ranking the energetic cost of (non-interacting) quantum error correcting codes. This opens possibilities for a variety of research avenues, and makes analyses of Landauer's principle  considerably more applicable to reversible and quantum computing.

\acknowledgements{It is a pleasure to thank Steve Campbell for many insightful discussions on the thermodynamics of quantum information, and Marcus V. S. Bonan\c{c} and Nathan M. Myers for helpful comments on the manuscript.  Further I am indebted to Tan Vu Van, who spotted a very unfortunate typo in the derivation. S.D. acknowledges support from the U.S. National Science Foundation under Grant No. DMR-2010127. }

\bibliographystyle{eplbib}
\bibliography{qubit_erasure_epl}

\end{document}